\documentstyle[preprint,prb,aps]{revtex}   

\begin{document}

\title{
Exchange coupling and current-perpendicular-to-plane giant
magneto-resistance of magnetic trilayers. Rigorous results within
a tight-binding single-band model.
}

\author{S.~Krompiewski and M.~Zwierzycki} 
 
\address{ Institute of Molecular Physics,
 P.A.N., Smoluchowskiego 17, 60-179 Pozna\'n, Poland}

\author{U.~Krey}

\address{ Institut f\"ur Physik II, Universit\"at Regensburg, 93040
 Regensburg, Germany}
          
\maketitle

\begin{abstract}
It is shown that the current-perpendicular-to-plane giant
magneto-resistance (CPP-GMR) oscillations, in the
ballistic regime, are strongly correlated with
those of the exchange coupling ($J$). Both  the GMR and
$J$ are treated
{\it on equal footing} within a rigorously solvable tight-binding
single-band model. The strong correlation consists in sharing 
asymptotically the same
period, determined by the spacer Fermi surface, and oscillating 
with varying spacer thickness predominantly in 
{\it opposite phases}. 
\end{abstract}
\vspace*{4mm}

The oscillatory behaviour of many physical phenomena of magnetic
multilayer systems manifests itself in the most spectacular way
as a function of 
spacer thickness, but the magnetic layer thickness is
relevant\cite{l:ba1,l:Euro,l:br}, too.
The most widely studied oscillatory
phenomena are those connected with either the exchange coupling
 ($J$) or the so called giant magneto-resistance (GMR)
(see Refs.\ 3 and 4 
for a review of  the currect
understanding of these
phenomena). The
exchange coupling  is of quantum nature and is well understood 
in terms of such theoretical approaches like: RKKY-type theory
{\cite{l:br}} , quantum well states {\cite{l:or}},
tight-binding model {\cite{l:ed}},
and free-electron-like one {\cite{l:ba2}}. From these approaches
as well as experimental results {\cite{l:ba}}  
and {\it ab initio} band structure
calculations\cite{l:he,l:Euro}, consensus
emerges on that the oscillation periods
of $J$ are determined by certain extremal spanning vectors of the
spacer Fermi surface. As regards the GMR, according to the
 two-spins channels
model, one expects a strong influence of the exchange coupling
(responsible for the mutual orientation of the magnetization of
 ferromagnetic slabs)
 on the resistivity. The anticipated trend would be 
to relate the antiparallel (parallel) orientation with maxima
 (minima) of GMR. The GMR can be easily measured if the relative
 spontaneous orientation of the magnetizations of the magnetic
slabs is
antiparallel (negative $J$), since then simply GMR$=(R(0)-R(H))/R(0)$,
where $H$ is the magnetic field necessary to switch to
the parallel orientation; but GMR remains well defined in the
opposite case, too. While the latter 
case makes no  problem for a theoretical treatment, it requires pretty
sophisticated handling (atomic engineering) in order to stabilize
the antiparallel orientation by pinning one of the ferromagnetic
slab magnetizations {\cite{l:di}}. 

Although the GMR in general is not of
quantum origin and contains some ingredients which are hard to control
(defects, impurities, surface and interface roughness $etc.$), there is
one contribution, due to reflections of electrons from quantum well
barriers, which is of the same origin as the exchange coupling.
 This quantum
contribution has been studied and shown to be quite substantial both
by {\it first  principles} computations {\cite{l:sche}}  and model
calculations\cite{l:ma,l:prb}.
The aim of the present paper is to confront the GMR oscillations in
the ballistic regime\cite{l:bu,l:ma,l:prb}, where only the quantum
 contribution appears, with those of the exchange
coupling. Both quantities are treated on equal footing without
any approximations, by precise numerical computations.

It is interesting that in spite of plenty of publications
on GMR and $J$, there has been,
to our knowledge, only one theoretical
attempt\cite{l:ba2}  devoted to a detailed comparison of both 
quantities. The authors of Ref.~7 
have used a free-electron-like
model, compared the $J$ behaviour with that of the
{\it current-in-plane} (CIP) GMR, and found that the CIP-GMR
assumes maxima for the {\it parallel} orientation.
That study could not treat, however, the relevant quantities on equal
footing since some inconsistency was unavoidable as a result of taking
into account electron scattering by impurities. As we will see below,
in our model the maxima, for the {\it current-perpendicular-to-plane}
 (CPP) GMR  arise for the {\it antiparallel} orientation 

In the present letter we put emphasis on selecting the above-mentioned
 CPP-GMR component of purely quantum origin, which can be
 directly compared
with the exchange coupling, and in principle, measured in the ballistic
regime {\cite{l:ho}}.
We adopt the rigorously solvable tight-binding single-band model
Hamiltonian 

\begin{equation}\label{eqn1}
H_{\sigma} = \sum \limits_{\vec i,\vec j}
t_{\vec i,\vec j} c_{\vec i,\sigma}^{\dag} c_{\vec j,\sigma} +
\sum_{\vec i} V_{\sigma}(\vec i) 
c_{\vec i,\sigma}^{\dag} c_{\vec i,\sigma}~,
\end{equation}
with $t$ being the nearest-neighbour hopping integral ($|t|$
is the energy
unit) and $V_{\sigma} (\vec i)$ -- the spin-dependent atomic potential.
The  systems under consideration are trilayers of the type
$n_f F/n_s S/n_f F$ , where $n_f$ $(n_s)$ stands for the number of the
ferromagnetic (non-magnetic spacer) monolayers in the perpendicular
$z$ -direction. The ferromagnetic slabs are magnetized either parallel 
or antiparallel to each other. Since the systems are infinite
in the $(x,y)$-plane and
the potentials $V_{\sigma}$ are only $z$-dependent, the 
eigenvalues of the Hamiltonian (\ref{eqn1}) for simple
cubic lattices
(with the lattice constant $=1$) are simply:

\begin{equation}\label{eqn2}
E(\vec{k}_{\|}, \tau )= \epsilon_{\bot}(\tau ) -2 
(\cos k_x + \cos k_y)\, , \end{equation}

where $ \epsilon_{\bot} $ are the eigen-values, labeled by $\tau$, of a 
tridigonal matrix of rank $2 n_f + n_s $ (with free boundary
conditions in the $z$-direction). The exchange splittings of
the ferromagnetic films have been introduced by taking
spin-dependent atomic potentials $V_{\sigma}$ outside the spacer. 
Additionally 
 we have assumed a
perfect matching of the minority bands in the whole system by putting
$V_{\downarrow} = 0$. The exchange coupling is then calculated in the
following way {\cite{l:ba1,l:sk}}

\begin{equation}\label{eqom}
\Omega = \sum \limits_{\vec {k}_{\|} ,\tau}
[E(\vec{k}_{\|} ,\tau ) - E_F ]\cdot
\theta (E_F - E(\vec{k}_{\|} ,\tau )),
\end{equation}

\begin{equation}\label{eqom1}
J = \Omega^{\uparrow\downarrow}_{\uparrow}  
+ \Omega^{\uparrow\downarrow}_{\downarrow}
- \Omega^{\uparrow\uparrow}_{\uparrow} 
- \Omega^{\uparrow\uparrow}_{\downarrow},
\end{equation}

i.e. $J$ is the difference of 
the grand-canonical thermodynamic potentials for
antiparallely and parallely magnetized configurations. The summation
in (\ref{eqom}) has been performed very accurately by means of the
{\it special k-points} method {\cite{l:cu}}.

The way we compute the CPP-GMR is the same exact one as in our  
recent paper {\cite{l:prb}}. It uses the Kubo formula and
applies an accurate recursion Green's function method to 
trilayer systems 
sandwiched between semi-infinite ideal lead wires. The method, based
on Refs.~18,19 
and modified by us in Ref.~13,
is rigorous. The GMR is defined by

\begin{equation} \label{eqn4}
{\rm GMR} = \frac{ \Gamma^{\uparrow\uparrow}_{\uparrow} +
 \Gamma^{\uparrow\uparrow}_{\downarrow}}{
 \Gamma^{\uparrow\downarrow}_{\uparrow} + 
\Gamma^{\uparrow\downarrow}_{\downarrow}}
-1 \, ,
\end{equation}

where $\Gamma$ is the conductance, and the superscripts and subscripts
indicate the  relative orientation of the magnetization, and the
 electron spin, respectively.

In Fig.~1 we present some typical plots of GMR $vs.$ $n_s$
for different
magnetic slab thicknesses ($n_f$). What is easy to see is that for
small $n_f$ the curves have a regular quasiperiodic behaviour which
seems to be disturbed for $n_f$ greater than, say, 6. Before we
explain this puzzling situation let us briefly remind that in the
asymptotic limit of large $n_s$ one expects a quasiperiodic behaviour
of the exchange coupling with a period depending exclusively
on the Fermi energy which (for a given lattice structure) determines
extremal spanning vectors of the spacer Fermi surface.
 Fig.~2 examplifies this for the $ 3F/10S/3F$
system with $E_F=2.5$. The Fermi surface has the form of 
$\epsilon_{\bot}$-constant contours with occupied electronic
states inside.
The arrow indicates the spanning vector $Q$ with the following 
components $[k_x=\pi,\, k_y=Q=0.774,\, \epsilon_{\bot}=1.929]$.
We quote the numerical values to show that already for a relatively
small system with 16 monolayers in the $z$-direction 
$\epsilon_{\bot}$ is close to the asymptotic value $2$, $i.e.$ 
$k_z =\pi$. Thus, asymptotically, for $n_s \to \infty$,
when the equivalence of all
the directions is restored and $k_z$ becomes a good quantum number,
one gets spanning vectors $[\pi, Q, \pi]$ and,
by symmetry, $[\pi, \pi, Q]$, $i.e.$ the oscillation
period $\lambda = \pi /Q \approx 4$.
Incidently, it is very easy to predict a period length in this limit,
by minimizing with respect to $k_x$ and $k_y$ the following function:

\begin{equation}\label{eqkz}
k_z(k_x,k_y,E_f)=\arccos ( -{E_f}/2 - \cos k_x -\cos k_y)\,,
\end{equation}

which gives $Q={k_z}^{extr} $ for $(k_x ,k_y ) = (0,0)$, $(\pm \pi ,0)$,
$(0, \pm \pi )$, $(\pm \pi ,\pm \pi)$. Another fact worth mentioning is
that the
influence of the magnetic slab thickness on the $J$  oscillations 
only leads to some phase shifts without actually
changing the period length\cite{l:br,l:sk}.

In Fig.~3 we show both GMR and  $J$ for $n_f = 3$ and
 $V_{\uparrow} = -1.8 $ and find pretty well correlated
oscillations with the same period consistent with the
strictly calculated
Fermi surface and the sketchy estimations above. It turns out that
this conclusion holds also for the other curves in Fig.~1, but to see it
one has to take a closer eye at them and allow for greater values
of $n_s$ to select the asymptotic trend. That has been made clear
in Figs.~4 ~and ~5 for $n_f = 7$ and $5$, with a period of about $10$
in the latter case. It is easily seen that the GMR does share with 
the exchange
coupling the long period of oscillations but has predominantly an
{\it opposite phase}\cite{l:rem},
 in the sense that for negative (positive) $J$ it takes
larger (smaller) values than its aymptotic value.  
This coincidence
is hardly visible for small $n_s$ until the asymptotic behaviour
develops, and is partially obscured by the superposition of some 
short period oscillations of GMR of non-RKKY nature
\cite{l:ma,l:prb}.

In conclusion, we have carried out  numerical studies of a rigorously
solvable tight-binding single-band model  -- treating both the CPP-GMR
 and
the exchange coupling  $J$ on equal footing  --  and found that
asymptotically both
quantities share the same oscillation period (consistent with
the Fermi surface topology) and have predominantly opposite phases.
Since according to Ref.~19  the CPP-GMR  is rather non-sensitive
to impurities, our statements should also
apply if additionally impurities are taken into account.

This work has been carried out under 
the bilateral project DFG/PAN 436 POL (U.K.,
 S.K.) and the KBN grants
No.~PB.872/PO3/96/11 and 2P 03B 165 10 (S.K.). We also thank
the  Pozna\'n, and Regensburg Computer Centres for computing time.

\newpage

{\Large \bf Figure Captions}
\vspace{1cm}

Fig.~1: CPP-GMR of the systems $ n_f F/n_s S/n_f F$ (where $n_f$
and $n_s$ are the number of ferromagnetic and non-magnetic spacer
monolayers, respectively).
Majority spin electrons have the potentials
$V_{\uparrow}=-1.8$ in the ferromagnets (all
other potentials are 0), $ E_F =2.5$.

\vskip 0.3 truecm 

Fig.~2: Fermi "surfaces" of the systems under consideration consist of
$\epsilon_{\bot}$-constant energy contours with occupied states inside,
where $\epsilon_{\bot}$ fulfils:  
$\epsilon_{\bot}-2(\cos k_x+\cos k_y)=E_F$.
The arrow indicates the extremal spanning vector $Q$ for the
$3F/10S/3F$ system (in the  {\it parallel} configuration,
with $E_F = 2.5$ and $V_{\uparrow}=-1.8$).

\vskip 0.3 truecm 

Fig.~3: GMR (solid line) and {\it exchange coupling (J)} (dashed
line) $vs.$ the spacer thickness, for $n_f=3$, $E_F=2.5$
and $V_{\uparrow}=-1.8$.

\vskip 0.3 truecm 

Fig.~4: The same as Fig.~3, but with $n_f=7$
and for higher values of $n_s$
to reveal the asymptotic trend of the oscillations.

\vskip 0.3 truecm

Fig.~5: The same as Fig.~4, but with $n_f=5$, $E_F=2.1$, and
$V_{\uparrow}=-2$.

\vskip 0.5 truecm 
\newpage 
\input epsf
\epsfxsize=18cm
\epsfbox{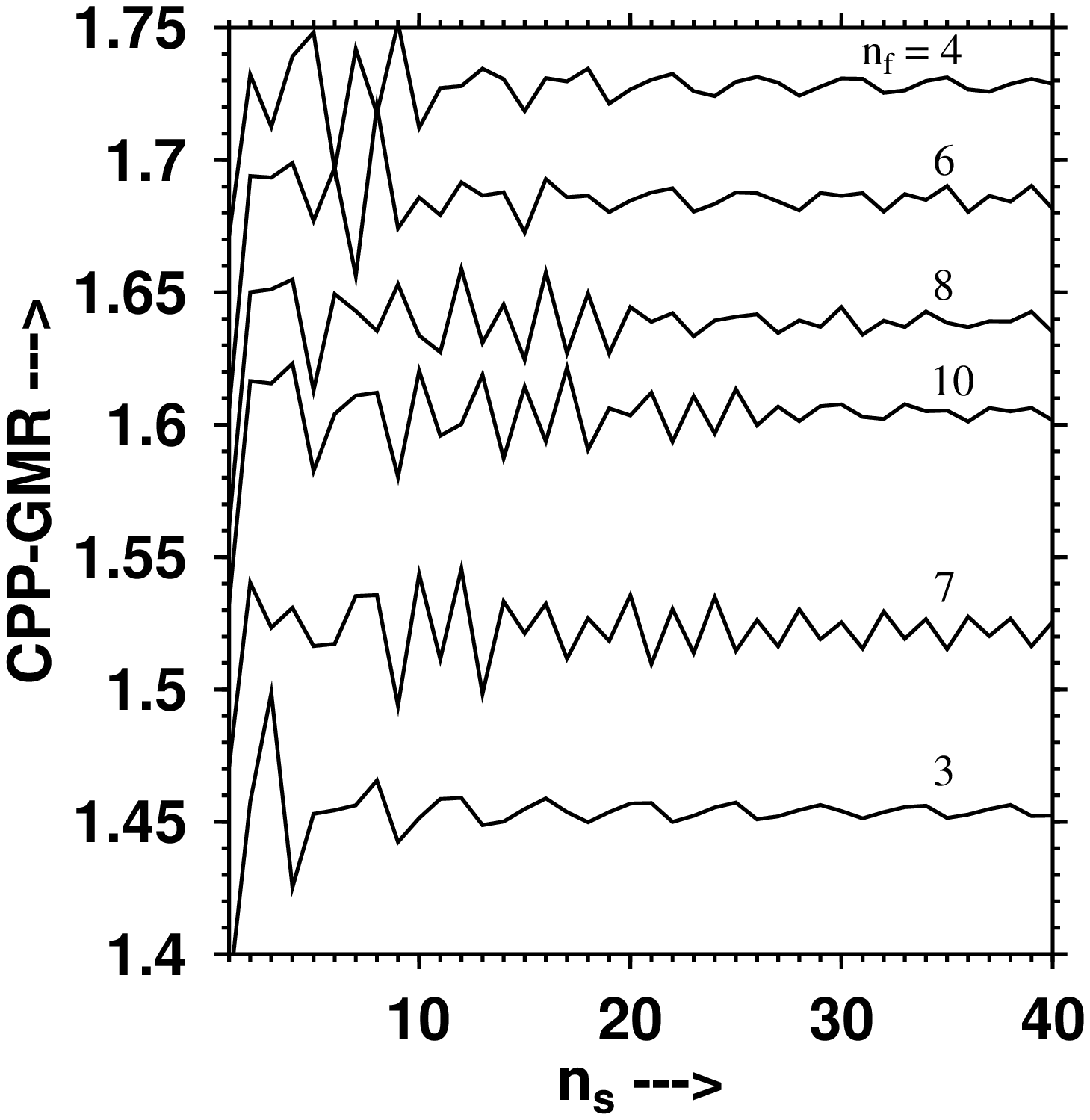} 
{\centerline{\underbar{Fig.~1}:}}
 \newpage
\newpage 
\epsfxsize=18cm
\epsfbox{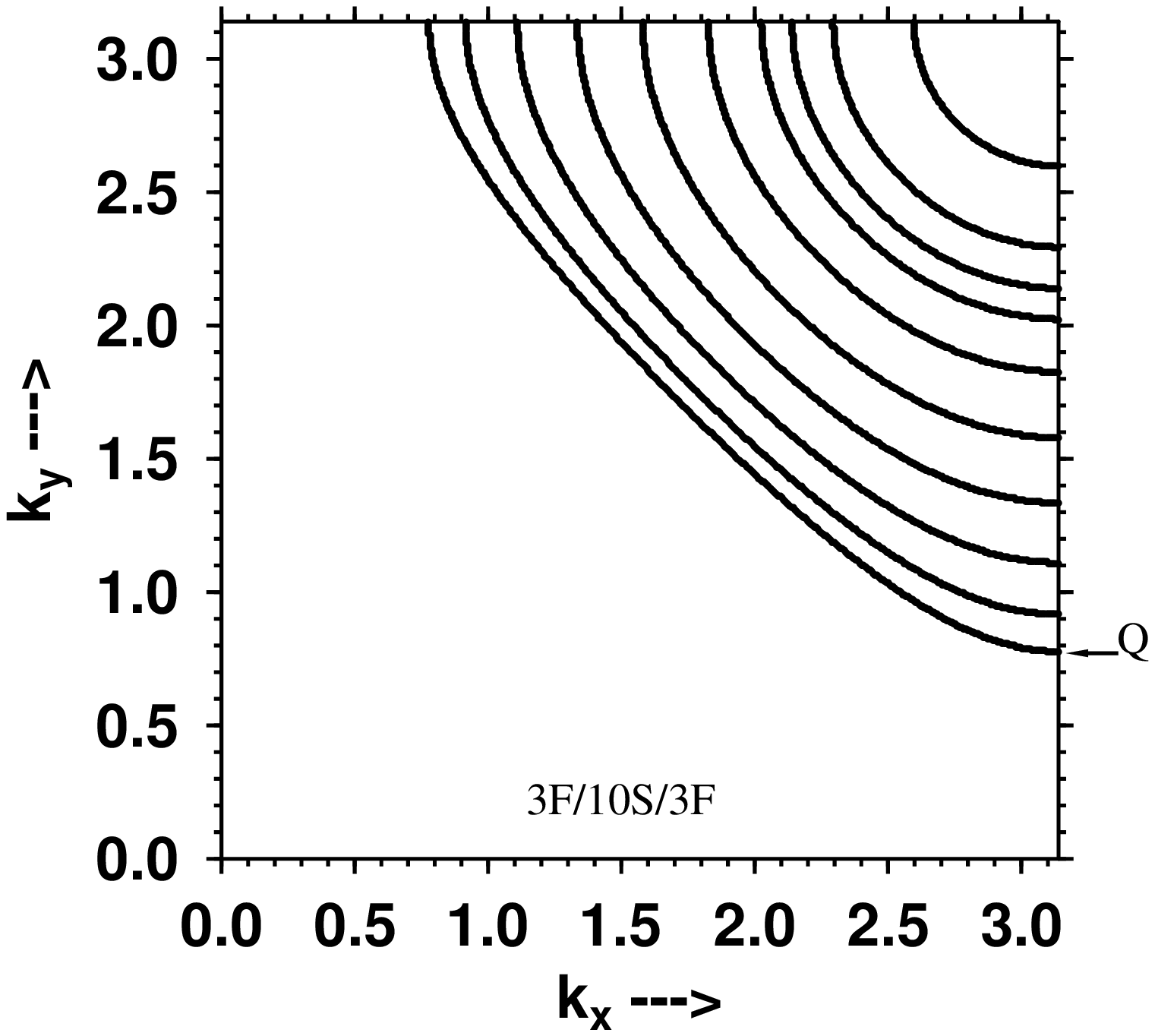} 
\centerline{\underbar{Fig.~2}:} 
 \newpage
\epsfxsize=18cm
\epsfbox{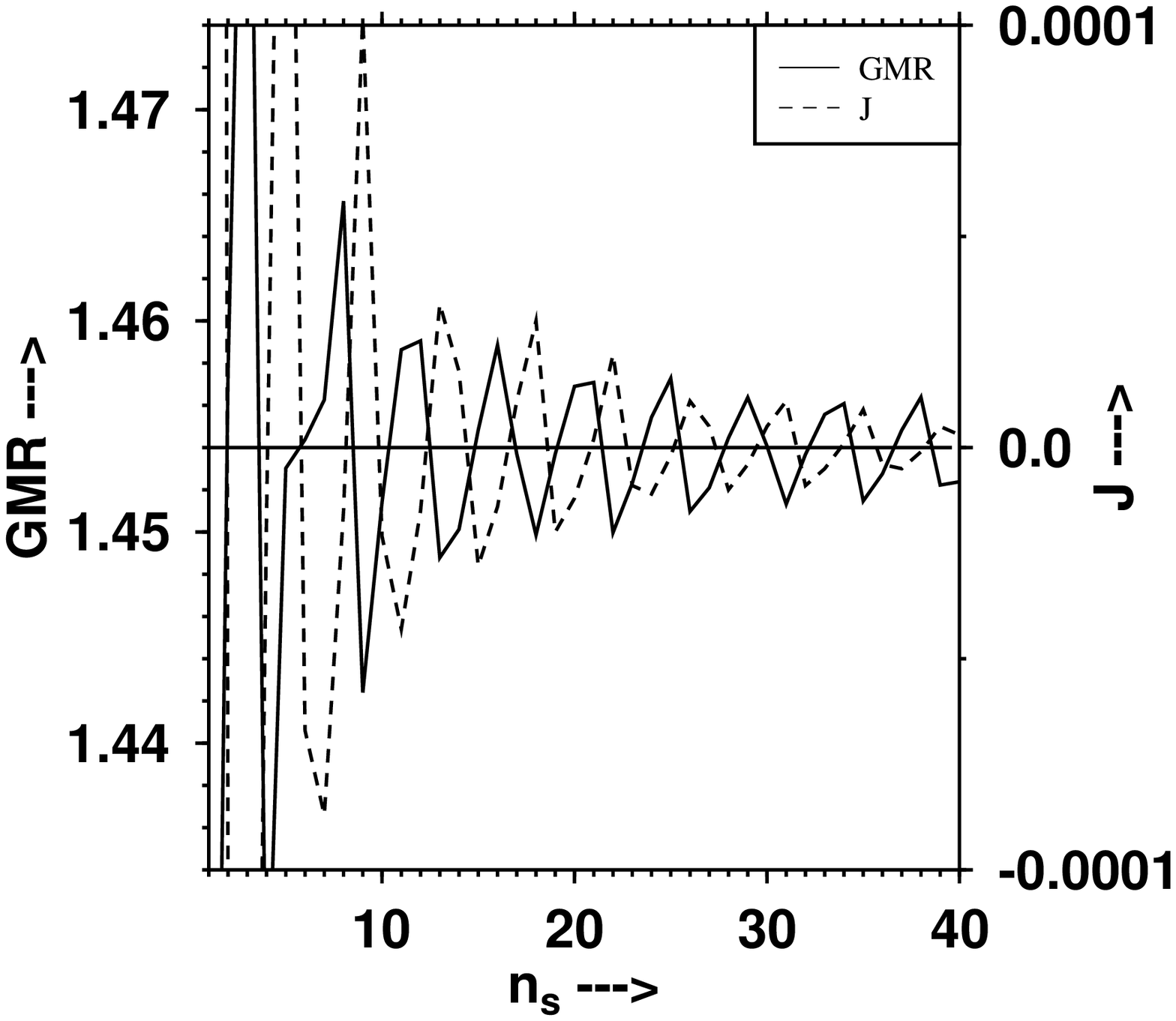} 
\centerline{\underbar{Fig.~3}:}
 \newpage
\epsfxsize=18cm
\epsfbox{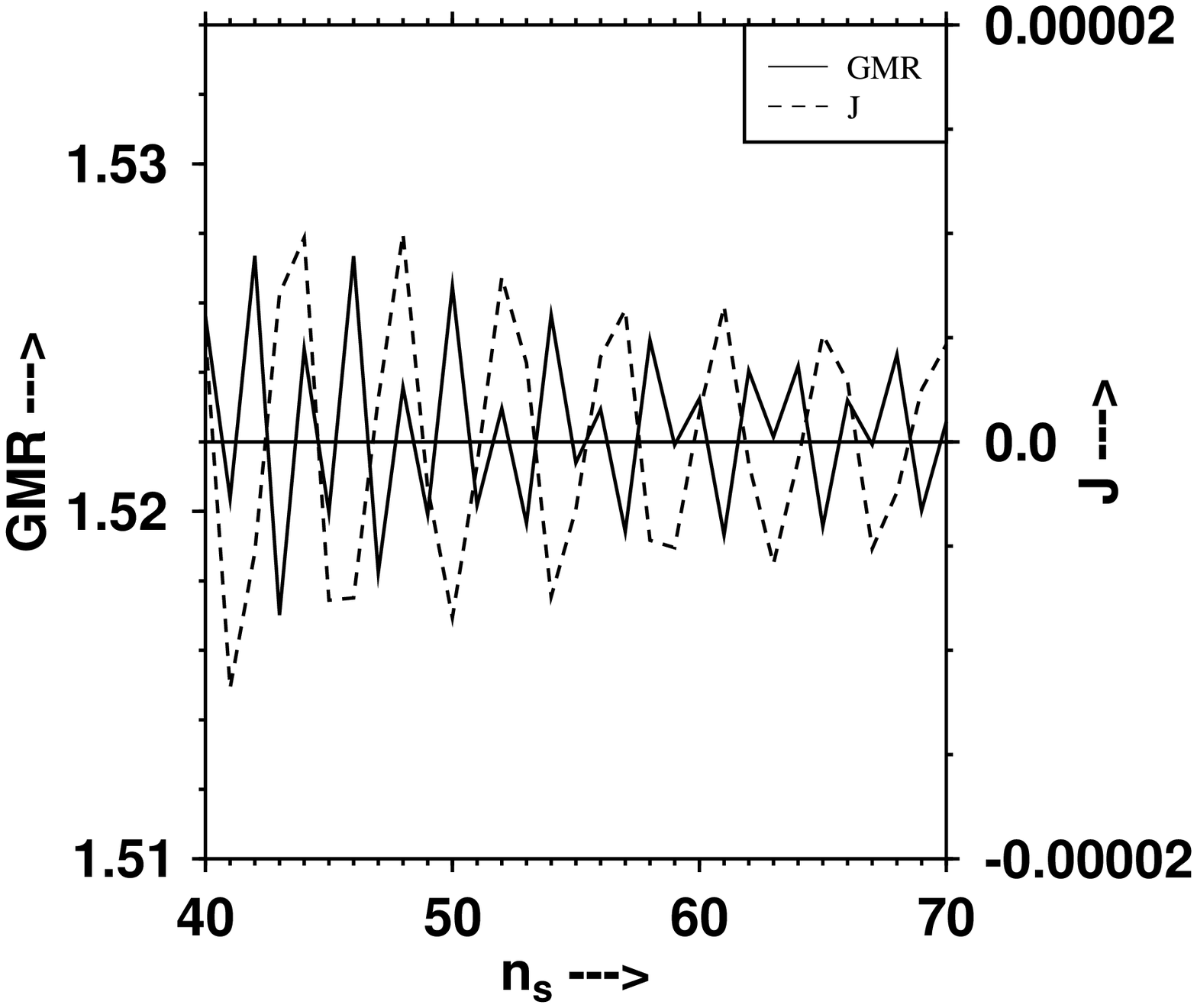} 
\centerline{\underbar{Fig.~4}:}
 \newpage
\epsfxsize=18cm
\epsfbox{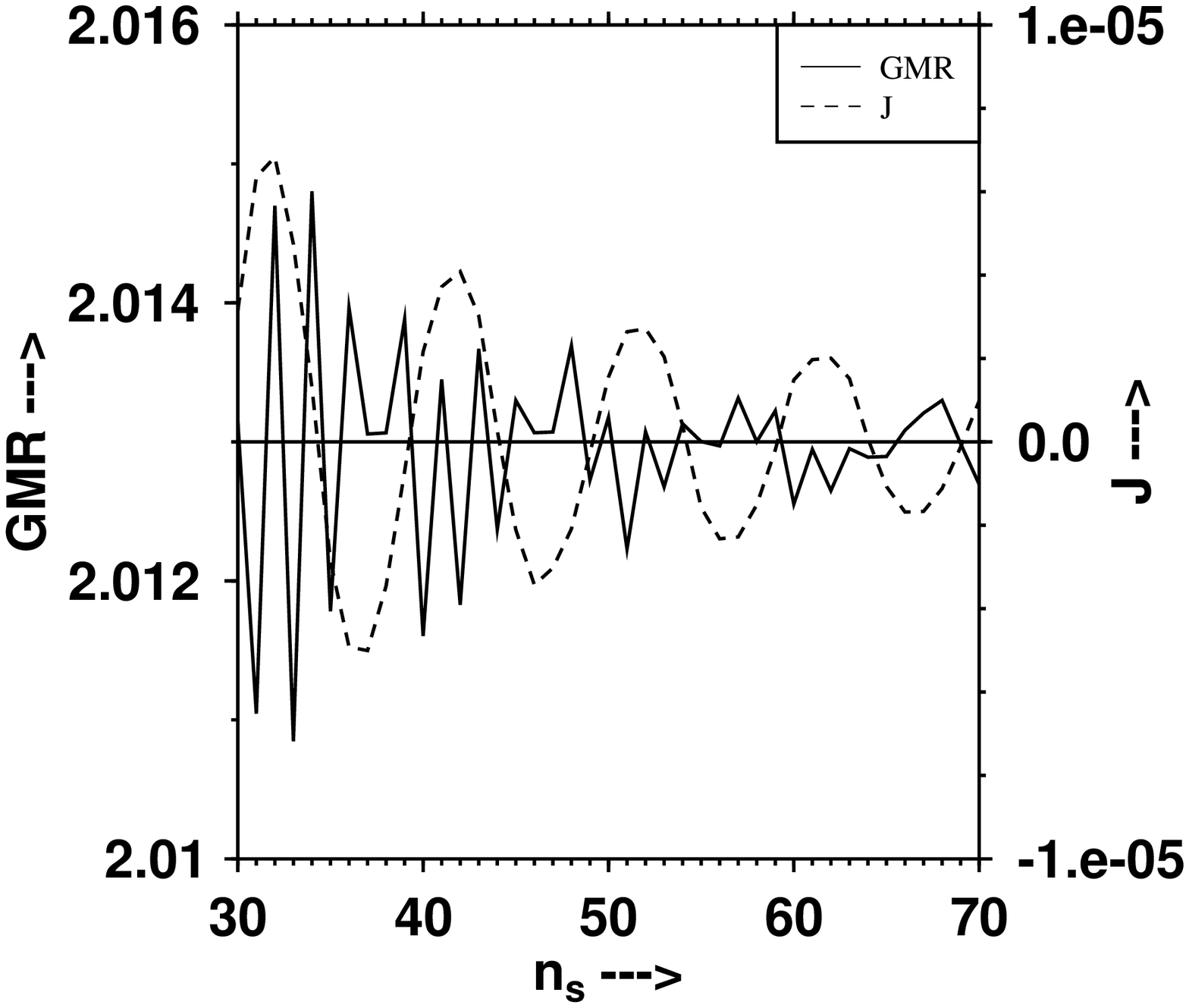} 
\centerline{\underbar{Fig.~5}:}
 \newpage
\end{document}